\documentclass[11pt]{article}

\usepackage[latin1]{inputenc}
\usepackage{amstext,amsmath,amssymb,amsfonts,bbm,amsmath}
\usepackage{extarrows}
\usepackage{xcolor} 
\usepackage[colorlinks, linkcolor=blue, citecolor=blue, urlcolor=blue]{hyperref}
\usepackage{authblk}
\usepackage{caption} 
\usepackage{epsfig} 
\usepackage{color} 
\usepackage{graphicx} 
\usepackage{braket} 
\usepackage{slashed} 
\usepackage{dsfont} 
\usepackage{amsthm}

\usepackage{cite}

\allowdisplaybreaks[4]

\usepackage{psfrag}
\usepackage{tikz} 
\usetikzlibrary{arrows}
\usetikzlibrary{plotmarks}
\usetikzlibrary{external}
\usetikzlibrary{patterns}
\usepackage{pgfplots}
\pgfplotsset{compat=1.9}
\usetikzlibrary{patterns}
\usetikzlibrary{calc}
\usetikzlibrary{automata,positioning}

\usepackage{float}  
\usepackage{subfig}
\usepackage[lmargin=50pt,rmargin=60pt,tmargin=60pt,bmargin=65pt]{geometry}

\DeclareMathOperator{\Tr}{Tr}

\newcommand{\be}{\begin{equation}}
	\newcommand{\ee}{\end{equation}} 

\newcommand{\f}{\frac}
\newcommand{\p}{\partial}

\let\a=\alpha \let\b=\beta  \let\g=\gamma    
\let\z=\zeta         
\let\m=\mu              \let\r=\rho 
\let\s=\sigma \let\t=\tau     
    
     \let\X=F
         
  \let\eps=\epsilon

\newcommand{\cL}{\mathcal{L}}

\newcommand{\cZ}{\mathcal{Z}}

\newcommand{\mbx}{\mathbf{x}}

\allowdisplaybreaks[4]

\numberwithin{equation}{section}

\theoremstyle{remark}

\begin{document}

	\title{\bf Dynamic critical exponent in quantum long-range models}

	\author[1]{Dario Benedetti}
 	\author[2]{Razvan Gurau}
  \author[2]{Davide Lettera}
	
\affil[1]{\normalsize\it 
CPHT, CNRS, \'Ecole polytechnique, Institut Polytechnique de Paris, 91120 Palaiseau, France
\authorcr \hfill }
	
	\affil[2]{\normalsize\it 
		Heidelberg University, Institut f\"ur Theoretische Physik, Philosophenweg 19, 69120 Heidelberg, Germany
		\authorcr \hfill
		}

	\date{}
	
	\maketitle
	
	\hrule\bigskip
	
	\begin{abstract}
    Quantum long-range models at zero temperature can be described by fractional Lifshitz field theories, that is, anisotropic models whose actions are short-range in time and long-range in space.
    In this paper we study the renormalization of fractional Lifshitz field theories with weakly relevant cubic or quartic self-interactions. Their nontrivial infrared fixed points exhibit Lifshitz scale invariance, and we compute the lowest-order corrections to the dynamic critical exponent.
	\end{abstract}
	
	\hrule\bigskip
	
	\tableofcontents
	

\section{Introduction}

Systems with long-range interactions appear in a large variety of physical situations \cite{Campa:2009rev}.
Among the many possible instances, a particularly interesting class of systems is the one described by models with a two-body interaction that decays with distance according to a power law, with sufficiently slow decay.
The paradigmatic example in such a class is the long-range Ising model, associated to the classical Hamiltonian
\begin{equation} \label{eq:Ising}
    H_I = -J \sum_{i,j\in\cL} \f{\s_i\,\s_j}{|i-j|^{d+2\z}} \ ,
\end{equation}
where $\cL \subset \mathbb{Z}^d$ is a finite $d$-dimensional lattice, $\s_i=\pm 1$ are the Ising variables at site $i\in\cL$, $J>0$ is their coupling, and the long-range exponent is in the range $0<\z<1$, with the limiting case $\z=1$ being equivalent to the nearest-neighbour (or short-range) model. 
Such model, and its extensions with internal $O(N)$ symmetry, has been extensively studied, both in its lattice version and in its field theoretic Ginzburg-Landau formulation \cite{Dyson:1968up,Fisher:1972zz,Sak:1973,Aizenman:1988critical,Brydges:2002wq,Abdesselam:2006qg,Slade:2017,Lohmann:2017,Brezin:2014,Paulos:2015jfa,Behan:2017emf,Giombi:2019enr,Benedetti:2020rrq,Behan:2023ile}, and several related experimental systems have been constructed in recent years  (see \cite{Defenu:2021glw} for a review).

We emphasize that $\cL$ in \eqref{eq:Ising} represents space only, time being frozen, as is always the case in classical equilibrium statistical mechanics.
When considering instead quantum versions of long-range models,
a short-range temporal interaction arises after the standard quantum-to-classical mapping \cite{Suzuki:1976cqdual,Sachdev:2011fcc}, as explicitly emphasized in \cite{Benedetti:2023pbt} (see also \cite{Dutta:2001,Defenu:2017}).\footnote{Models with long-range temporal dynamics can also be conceived, for example in the presence of dissipation, impurities, or memory effects (e.g.\ \cite{Winter_2009,Zeng:2023temporal}), but we will not consider them here.}
Keeping with the Ising example, the quantum model is obtained by replacing the classical Ising variables $\s_i$ with Pauli matrices $\hat{\s}_i^z$ at each site, and adding to the Hamiltonian a transverse field interaction $h\sum_i \hat{\s}_i^x$, in order to induce a nontrivial quantum dynamics:
\begin{equation}
    \hat{H}_{qI} = -J \sum_{i,j\in\cL} \f{\hat{\s}_i^z\,\hat{\s}_j^z}{|i-j|^{d+2\z}} - \g \sum_{i\in\cL} \hat{\s}_i^x \;.
\end{equation}
The quantum statistical partition function at inverse temperature $\beta$ can then be mapped to a classical one by applying the Trotter formula to $e^{-\b \hat{H}_{qI}}$, and after some manipulations \cite{Suzuki:1976cqdual} obtain:
\begin{equation} \label{eq:ZqI}
    \cZ_{qI} = \Tr[e^{-\b\hat{H}_{qI}}] 
    \sim \lim_{n\to\infty} \sum_{\{\s_{i,t}=\pm 1\}_{i\in\cL,t=1\ldots n}} \exp\Bigg\{\f{\b J}{n} \sum_{i,j,t} \f{\s_{i,t}\,\s_{j,t}}{|i-j|^{d+2\z}} + \f{\b}{2}\ln\coth(\f{\g}{n}) \sum_{i,t} \s_{i,t}\s_{i,t+1} \Bigg\} \;,
\end{equation}
where the classical Ising variables now have a second index, interpreted as discrete Euclidean time, and periodic boundary conditions are assumed: $\s_{i,n+1}=\s_{i,1}$.
One then concludes that the Ginzburg-Landau description of the quantum long-range Ising model must be an anisotropic scalar field theory, long-range in space and short-range in time.
In \cite{Benedetti:2023pbt} we called such a theory a  fractional Lifshitz field theory, and the scope of this paper is to further study its critical properties.

Scalar Lifshitz field theories were first studied in \cite{Hornreich:1975zz} to describe tri-critical points in the presence of paramagnetic, ferromagnetic, and modulated phases. 
Denoting the time coordinate $\tau$ and the $d$-dimensional spatial coordinates $x$, a standard example is an action of the form:\footnote{From now on, when discussing the field-theoretic approach to these models we will talk about actions rather than Hamiltonians.}
\begin{equation}\label{ExampleAnisAction}
    S[\phi]= \int d^dx d\tau\Bigg\{\frac{1}{2} (\partial_{\tau}\phi)^2+\frac{\rho_0}{2} \sum_j (\partial_{j}\phi)^2+\frac{1}{2} \Big(\sum_j \partial_j \partial_{j}\phi\Big)^2+V[\phi]\Bigg\} \ ,
\end{equation}
with $j$ ranging over the spatial dimensions. 
Usual ordered and disordered phases can be reached by tuning the potential, while a spatially modulated phase, breaking translation invariance, can be obtained by choosing $\r_0<0$. The Lifshitz point \cite{Hornreich:1975zz} is by the definition the critical point at the intersection of such phases, and one is interested in computing the associated critical exponents.
For example, this has been done for a model in this family with a quartic interaction, using either perturbative methods up to two loops \cite{Diehl:2000sv,SHPOT2001340,Diehl:2002ri} or the functional renormalization group approach \cite{Essafi:2012hr}, while in \cite{Shpot:2004gc,Shpot:2008gfh,Shpot:2012mw} the anisotropic quartic $O(N)$ vector model was studied by means of the $1/N$ expansion. 
Notice that in the original setting of Lifshitz points, ``time" is actually one spatial direction of a spatially anisotropic classical system, but it can have an interpretation as (real or imaginary) time in other contexts, such as Lorentz-violating quantum field theories with improved ultraviolet behavior \cite{Anselmi:2007ri,Horava:2009uw,Zappala:2021hzm}, or quantum Lifshitz points \cite{Ardonne:2003wa}. 

One interesting feature of Lifshitz field theories is that while they break explicitly Lorentz or rotation invariance, they exhibit anisotropic scale invariance at fixed points of the renormalization group. 
An  anisotropic scaling transformation of an operator ${\cal O}$ writes:
\begin{equation} \label{AnisScaleInvariance}
    \mathcal{O}(\tau , x)=l^{ \Delta_\mathcal{O}} \mathcal{O}(l^z \tau,l x) \ ,
\end{equation}
with $\Delta_\mathcal{O}$ the scaling dimension of ${\cal O}$, and $z\neq 1$ the anisotropy exponent, also known as dynamic critical exponent due to its similar role in dynamic critical phenomena \cite{Hohenberg:1977ym}. 
Typical Lifshitz field theories in the literature are constructed as perturbations of a Gaussian theory with an integer anisotropy exponent, as in the example above, which has $z=2$. However, a non-integer $z$ generally appears at interacting fixed points.
The way this happens in practice is that in the renormalized theory the operators $(\partial_{\tau}\phi)^2$ and $(\sum_j \partial_j \partial_{j}\phi)^2$ get different anomalous dimensions, say $\eta_2$ and $\eta_4$, respectively, and one finds the relation \cite{Diehl:2002ri,Essafi:2012hr}:
\begin{equation}
    z = \f{4-\eta_4}{2-\eta_2} \;.
\end{equation}

Here, we will study a Lifshitz type of model in which a nonlocal operator, known as fractional Laplacian, replaces the higher-order spatial derivative terms.\footnote{From the point of view of our motivation the time direction has the interpretation of imaginary time in the classical description of a quantum statistical model. However, a spatial interpretation of an anisotropic classical statistical model is also possible (see for example \cite{Defenu:2016pwy} for a more general model with this interpretation).} In other words, we will consider a continuum version of the model \eqref{eq:ZqI}, with the long-range kernel corresponding to the integral representation of a Laplacian to power $\zeta$, with $0<\zeta<1$.  
As a consequence of that, the standard short-range term $(\partial_{j}\phi)^2$ is in this case irrelevant,\footnote{However, beyond some value of the long-range exponent it might become a dangerously irrelevant operator, as in the isotropic case \cite{Sak:1973,Behan:2017emf}.} and thus there is no modulated phase in these models.
Moreover, the nonlocal nature of the spatial term implies that it needs no renormalization, and hence its anomalous dimension vanishes.
Therefore, we expect to find $z=2\z/(2-\eta_2)$, which we will confirm.

We will consider models with either a cubic or a quartic interaction in the regime in which they are weakly relevant, that is, for $\zeta$ slightly above its lower critical value (i.e.\ the value below which the infrared fixed point is non-interacting, and thus mean field theory applies).
We will show that they exhibit a nontrivial infrared renormalization group fixed point, and we will compute  explicitly the corrections to the canonical value of $z$ (i.e.\ $z=\z$) at leading order in the perturbative expansion.

The model with quartic interaction has $\mathbb{Z}_2$ invariance, and it can be interpreted as the Ginzburg-Landau theory for the quantum long-range Ising model, as explained above.
The cubic model instead will require an imaginary coupling, and thus it can be interpreted as describing the Yang-Lee edge singularity at imaginary magnetic field for the same model, similarly to the usual Yang-Lee model \cite{Fisher:1978pf} (recently reviewed in \cite{Cardy:2023lha}).

We will start from the cubic model in Section~\ref{sec:cubic}, as calculations in this case are slightly easier, and as such it provides a useful benchmark for general renormalization aspects. Moreover, such model has never been studied before. 
We will then move to the quartic case in Section~\ref{sec:quartic}. The latter has been studied also in \cite{Dutta:2001,Defenu:2017}, but with methods and results that are complementary to ours.

\section{The cubic model}
\label{sec:cubic}

\paragraph{The free model.} The free fractional Lifshitz theory in $d+1$ dimensions with Euclidean signature is defined by the action:
\begin{equation}
    S[\phi] = \frac{1}{2} \int d \tau d^d x \; \phi(\tau,x)
   \left[  \left(-\partial^2\right)^\z - \partial_{\tau}^2\right]\phi(\tau,x)   \; ,
\end{equation}
where $\tau$ is the Euclidean time coordinate, $x$ the $d$-dimensional position, and $(-\partial^2)^\z=\left(-\sum_{i=1}^d \partial_{i}\partial^i\right)^\zeta$ stands for the fractional Laplacian in the $d$ spatial dimensions \cite{Kwasnicki:2017}. 
In position space, the latter is given by a nonlocal integral kernel which corresponds to the continuum version of \eqref{eq:Ising}:
\begin{equation}
   \int  d^d x \; \phi(\tau,x) \left(-\partial^2\right)^\z \phi(\tau,x) \propto \int  d^d x \, d^d y \; \f{\phi(\tau,x)\phi(\tau,y)}{|x-y|^{d+2\zeta}} \; .
\end{equation}
For the sake of conciseness, and for stressing the analogy to ordinary Lifshitz field theories, in the rest of the paper we will use the derivative notation when writing the action.

The covariance of the model, written as Fourier transform, is: 
\begin{equation}
C(\tau,x) = \int \frac{d^dp \; d\omega}{(2\pi)^{d+1} } \; 
\frac{e^{\imath \omega \tau + \imath p x}}{  \omega^2 + (p^2)^\zeta} \; , 
\end{equation}
and it enjoys the anisotropic Lifshitz scale invariance:
\begin{equation}
    C(\tau,x) = l^{2\Delta_\phi} \, C(l^{\zeta} \tau ,l x) \;,
\end{equation}
with $\Delta_\phi = \frac{d-\zeta}{2}$ the mass dimension of the field under anisotropic scaling. As the model is free, the correlators are computed by the Wick theorem and obey the scaling law:
\begin{equation}
F^{(n)}(\tau_1,  x_1 ; \dots ;  \tau _{n}, x_{n} ) 
 = \frac{\int [d\phi] e^{-S(\phi) } \; \phi( \tau_1,x_1) \dots
 \phi( \tau_{n},x_{n} )  }{ \int [d\phi] e^{-S(\phi) } }
 = l^{n\Delta_{\phi}} \; F^{(n)}(l^{\zeta} \tau_1,l x_1 ; \dots ; l^{\zeta} \tau_{n},l x_{n} ) \;,
\end{equation}
reflecting the invariance of the action under a field change of variable
$\phi(\tau,x) = l^{\Delta_\phi} \phi'(l^\z \tau,lx)$,
where we took into account the fact that the Jacobian of this change of variable  is field independent. 

Only the correlators with even $n$ are non zero, and if one restricts to connected correlation functions $G^{(n)}$, only the two point function is nontrivial and it equals the covariance:
\begin{equation}
G^{(2)}(\tau_1,x_1 ; \tau_2,x_2) = C(\tau_1- \tau_2, x_1-x_2) \;.
\end{equation}

\paragraph{The interacting model: bare theory and scaling.}
For the interacting theory one needs to distinguish between the bare and the renormalized versions of the model. We start by discussing the bare theory, defined by the action:
\begin{equation}
    S_b[\phi_b] = \int d \tau d^d x \Bigg\{ \frac{1}{2}\phi_b(\tau,x)
   \left[  \left(-\partial^2\right)^\z - \partial_{\tau}^2\right]\phi_b(\tau,x)+\imath \frac{\lambda_b}{3!} \phi_b(\tau,x)^3  \Bigg\} \; ,
\end{equation}
which depends on the bare field $\phi_b(\tau,x)$. Depending on the regularization scheme, additional linear and quadratic terms might be needed for renormalization, but we can view the above action as representing our target scale-invariant theory.
No couplings are introduced in the kinetic term, as they can always be set to 1 by a rescaling of $\phi_b$, $\t$, and $\lambda_b$. Note that we have explicitly factored an imaginary unit in the interaction, which is standard for cubic interactions.
We will also assume $\lambda_b>0$, since the two choices of sign are related by a field redefinition $\phi_b\to-\phi_b$.

The critical value of $\z$ for which the cubic interaction is marginal, that is such that the interaction is invariant under the field change of variable $\phi_b(\tau,x) = l^{\Delta_\phi} \phi'_b(l^\z \tau,lx)  $, respects $3\Delta_\phi - d - \zeta  =0$, that is $\z=\frac{d}{5}$. The weakly relevant case is obtained by setting $\zeta=\frac{d+\eps}{5}$, with $\epsilon>0$ small, leading to the mass dimension of the coupling $[\lambda_b]= d+\zeta -3\Delta_\phi=\frac{\eps}{2}$. 
In order to have an interesting renormalization group flow towards the infrared, and keep $\z<1$, which is needed for the long-range interpretation and for avoiding the need of a relevant $\phi_b \p_x^2\phi_b$ term in the action, we must stick to $d<5$.

With the above choices, the action is invariant under the simultaneous change of field variables and coupling:
\begin{equation}
  \phi_b(\tau, x) =   l^{\Delta_\phi} \phi'_b(l^\z \tau,lx)  \;, \qquad \lambda_b = l^{  \eps/2 } \lambda_b'  \qquad \Rightarrow \qquad S_b[\phi_b,\lambda_b] = S_b[\phi'_b,\lambda'_b] \; .
\end{equation}
We stress that this is essentially dimensional analysis, not a a true invariance of the theory, because 
the actions above are evaluated at different values of the coupling.

For $\eps>0$, using analytical continuation when needed, the correlation functions of the theory are finite order by order in the perturbative expansion, 
and they display the following behaviour under rescaling (we denote the arguments of the correlators collectively by $\tau,x$):
\begin{equation} \label{ScaleTrasfGn}
F^{(n)}_b(\tau,x | \lambda_b) = 
 \frac{\int[d\phi_b] e^{-S_b[\phi_b,\lambda_b] } \; \phi_b(\tau_1,x_1) \dots \phi(\tau_n,x_n) }{ \int[d\phi_b] e^{-S_b[\phi_b,\lambda_b] }  } 
= l^{n\Delta_\phi} F_b^{(n)}(l^\zeta \tau, l x | l^{- \eps/2 }\lambda_b ) \; .
\end{equation}
Equivalently, the bare correlation functions respect the differential equation:
\begin{equation}
\bigg[ n\Delta_\phi  - \frac{\epsilon}{2} \lambda_b \partial_{\lambda_b} + \zeta D_\tau + D_x \bigg] F_b^{(n)}(\tau,x | \lambda_b ) =0 \;,
\end{equation}
where $D_\tau = \sum_i \tau_i \partial_{\tau_i}$ is the time and $D_x = \sum_{i,\nu} x_i^\nu \partial_{x_i^\nu} $ the space dilatation operator. Formally, the bare correlators are eigenfunctions of the anisotropic dilatation operator $\zeta D_\tau + D_x$ in the marginal $\epsilon=0$ case, and slightly break it in the weakly relevant case. 

This is however formal because, 
as usual, the bare correlation functions display poles in $1/\epsilon$ which we will eliminate by passing to renormalized ones.

\paragraph{The effective action.} 
One often considers the effective action of the model, that is minus the generating function of amputated one-particle irreducible (1PI) correlators. For the bare theory this writes (somewhat formally):
\begin{equation}
 e^{-\Gamma_b[\Phi_b]} = \int_{\rm 1PI} d\varphi \; e^{-S_b[\Phi_b +\varphi]} \; .
\end{equation}

The transformation of the bare action under scaling implies that:
\begin{equation}
\Gamma_b[\Phi_b | \lambda_b]\Big{|}_{\Phi_b = l^{\Delta_\phi} \Phi'_b(l^\zeta \tau,l x) ; \lambda_b = l^{\eps/2} \lambda_b'}
= \Gamma_b[\Phi'_b | \lambda_b'   ]  \; ,
\end{equation}
and it follows that under a rescaling the bare amputated 1PI correlators behave as:\footnote{
We use
$  \int d\tau dx \; \Phi_b(\tau,x)^n \Gamma_b^{(n)}(\tau,x| \lambda_b)
     = \int d\tau' dx' \left( l^{-\Delta_\phi} \Phi_b(l^{-\z} \tau',l^{-1} x')  \right)^n \Gamma_b^{(n)}(\tau',x'|l^{- \eps/2 } \lambda_b) $.
}
\begin{align}\label{eq:gammabareresc}
& \Gamma_b^{(n)}(\tau,x | \lambda_b ) =  l^{n (\zeta + d - \Delta_\phi)} \Gamma_b^{(n)}(l^\zeta \tau, l x | l^{- \eps/2 }\lambda_b )   \;, \crcr
&\bigg[ n(\zeta + d -\Delta_\phi)  - \frac{\epsilon}{2} \lambda_b \partial_{\lambda_b} + \zeta D_\tau + D_x \bigg] \Gamma_b^{(n)}(\tau,x | \lambda_b ) =0 \;.
\end{align}

In momentum space the scaling transformation becomes
$\Gamma_b^{(n)} (\omega,p |\lambda_b) =  l^{ - n \Delta_\phi} \Gamma_b^{(n)}(l^{-\z} \omega, l^{-1} p | l^{- \eps/2 }\lambda_b ) $, but we note that there is one subtlety: this scaling relation concerns the \emph{full} $n$-point amputated correlators.  Such correlators contain a global conservation of momentum and frequency:
\begin{align}
& \Gamma_b^{(n)} (\omega,p |\lambda_b) = \delta(\sum \omega)
\delta(\sum p)  \; \bar  \Gamma_b^{(n)} (\omega,p |\lambda_b) 
\;, \crcr
& \bar \Gamma_b^{(n)} (\omega,p |\lambda_b) = l^{\zeta+d} l^{ - n \Delta_\phi} \bar \Gamma_b^{(n)}(l^{-\z} \omega, l^{-1} p | l^{- \eps/2 }\lambda_b )  \; ,
\end{align}
where we denoted $\bar \Gamma^{(n)}$ the momentum space $n$-point kernels of the effective action.\footnote{In order to underline the distinction between $\Gamma^{(n)}$ and $\bar \Gamma^{(n)}$, observe that for an isotropic long range free theory we have the two point amputated correlator $\Gamma^{(2)}(p,q) = p^{2\z}\delta(p+q)$, $d = 2\z + 2\Delta_\phi$ which obeys the scaling law $ l^{-2\Delta_\phi} [ l^{-2\zeta} p^{2\zeta} \delta(l^{-1} p + l^{-1} q) ] = p^{2\z} \delta(p+q) $, while $ \bar \Gamma^{(2)}(p) = p^{2\zeta} $ obeys the scaling law $l^{d-2\Delta_\phi} (l^{-2\z} p^{2\z})=p^{2\z}$.}

\begin{figure}[htb!]
    \centering
    \includegraphics[width=0.5\textwidth]{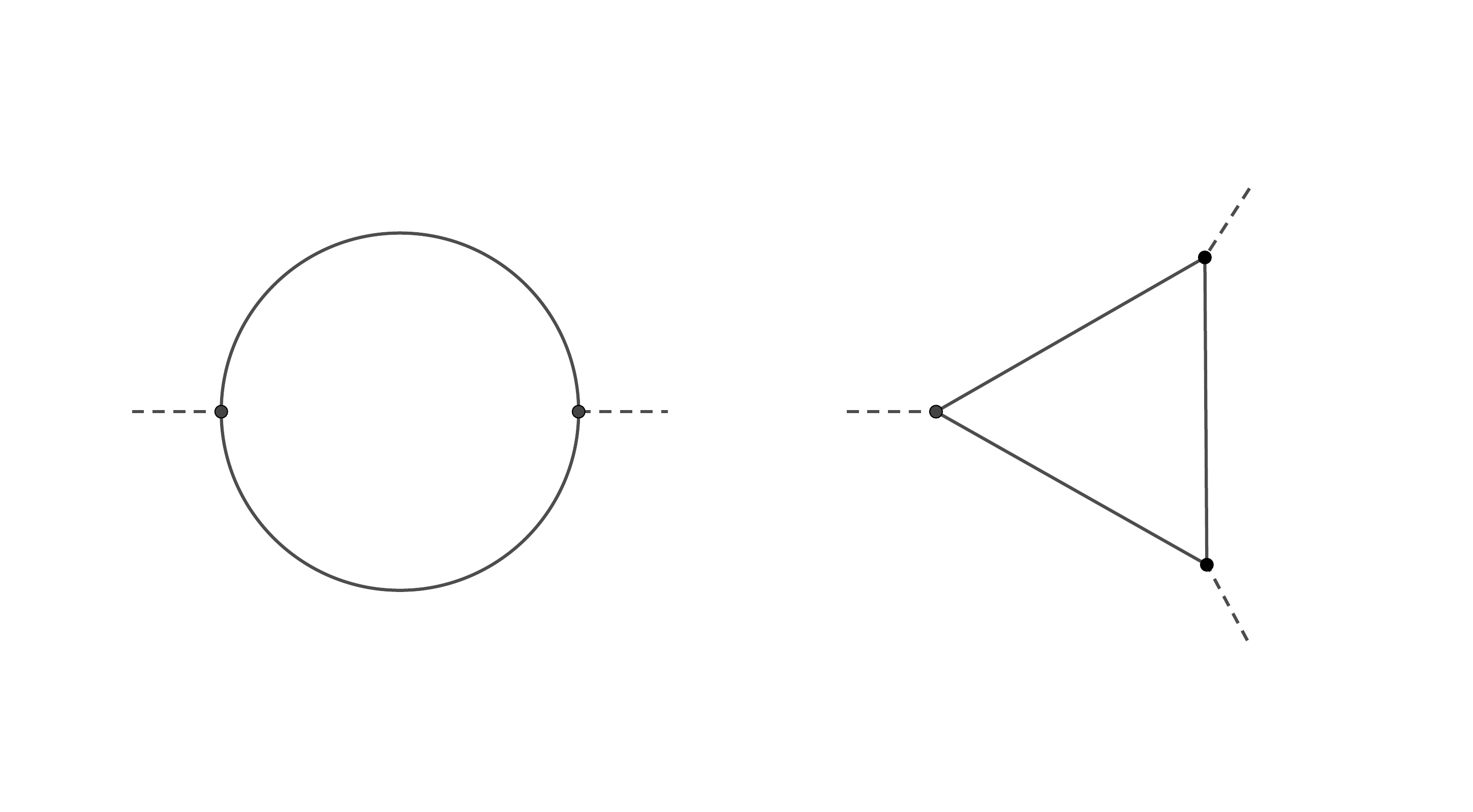}
    \caption{Left: bubble diagram $\mathcal{B}(\omega_0,p_0)$. Right: triangle diagram $\mathcal{T}(\omega_1,p_1;\omega_2,p_2)$.}
    \label{fig:BubbleTriangle}
\end{figure}

\paragraph{The one loop order.}
At one loop order, denoting $\mbx = (\tau,x)$ and $d\mbx = d\tau d^d x$, the bare effective action is expanded in powers of the field as follows: 
\begin{equation}
\begin{split}
\Gamma_b[\Phi_b] = & S_b[\Phi_b]  + \imath \frac{\lambda_b}{2}   \; \int d\mbx \; \Phi_b(\mbx) C(0) 
+ \frac{\lambda_b^2}{4}    \; \int d\mbx d\mbx' \; \Phi_b(\mbx) \Phi_b(\mbx' )  C(\mbx -\mbx')^2 \crcr
&  - \imath \frac{\lambda_b^3}{3!}  \; \int d\mbx d\mbx' d\mbx'' \; \Phi_b(\mbx ) \Phi_b(\mbx')  \Phi_b(\mbx'')  C( \mbx -\mbx' )C(\mbx' -\mbx'' )
C(\mbx'' -\mbx)  + O(\Phi^4) \; .
\end{split}
\end{equation}
The 1PI bare two and three point functions in momentum space, where we factored out the global momentum and frequency conservation, are at this order:
\begin{equation}
\bar \Gamma_b^{(2)}(\omega_0,p_0)  =\omega_0^2+ (p_0^2)^{\z}+\frac{1}{2}\lambda_b^2 \ \mathcal{B}(\omega_0,p_0)\ ,  \qquad
\bar \Gamma_b^{(3)}(\omega_1,p_1;\omega_2,p_2) = \imath \lambda_b - \imath \lambda_b^3 \ \mathcal{T}(\omega_1,p_1;\omega_2,p_2) \ ,
\end{equation}
where $\mathcal{B}$ and $\mathcal{T}$ are respectively the bubble and the triangle diagrams depicted in Fig.~\ref{fig:BubbleTriangle}
with amplitudes:
\begin{align}
&  \mathcal{B}(\omega_0,p_0) = \int \frac{d^dq \; d\omega}{(2\pi)^{d+1} }  
\;\frac{1}{ \omega^2 + (q^2)^\zeta } \; \frac{1}{ (\omega_0+\omega)^2 + 
[(p_0 + q)^2 ]^\zeta } \ , \crcr
& \mathcal{T}(\omega_1,p_1;\omega_2,p_2) = \crcr
& \qquad =\int  \frac{d^dq \; d\omega}{(2\pi)^{d+1} }  \; 
\frac{1}{ \omega^2 + (q^2)^{\zeta} } \; \frac{1}{(\omega_1+\omega)^2 + 
[(p_1 + q)^2 ]^\zeta } \; \frac{1}{(\omega_1 + \omega_2+\omega)^2 + 
[(p_1 + p_2 + q)^2 ]^\zeta } \; ,
\end{align}
where we have labelled all the momenta as incoming.

\paragraph{Power counting.}
At arbitrary orders, the effective action writes as the sum over 1PI amputated Feynman graphs with amplitude:
\begin{equation}
{\cal A}(\omega_E,q_E) = \int \prod_{L} d\omega_L d^dq_L \; \prod_{e} \frac{1}{  \omega_e^2 + (q_e^2)^\z } \; ,
\end{equation}
where $q_L,\omega_L$ are the independent loop momenta and frequencies, $q_e, \omega_e$ are the momenta and frequencies of the internal edges, that is linear combinations of loop momenta and frequencies and external momenta and frequencies $q_E,\omega_E$. Note that we have factored a global conservation of frequency and momentum.

The power counting of a graph with $P$ propagators (edges), $V$ vertices and $n$ external points is obtained by taking $q_L\sim \Lambda$ and $\omega_L\sim \Lambda^\zeta$, leading to $\Lambda^{(d+\zeta)(P-V+1) - 2\zeta P}$. Taking into account that $2P=3V-n$, this is:
\begin{equation}
  \Lambda^{ d+\zeta - (d+\zeta)V + (d-\zeta )P } =
  \Lambda^{d+\zeta - \frac{n}{2} (d-\zeta) + V (\frac{d}{2} -\frac{5}{2} \zeta ) } \;,
\end{equation}
which for $\zeta = \frac{d}{5}$ becomes $\Lambda^{\frac{d}{5} (6 - 2n)}$. We conclude that, at $\eps=0$:
\begin{itemize}
\item vacuum and one-point graphs ($n=0,1$) are power divergent: the vacuum graphs play no role and the one-point graphs are not 1PI, except for a generalized amputated tadpole. A linear counterterm will be added in order to cancel this term and ensure that $\Phi=0$ is a stationary point for the effective action.

\item two-point graphs ($n=2$) are power divergent, ${\cal A}(\omega_0)\sim \Lambda^{\frac{2d}{5}}  $, and $\partial_{\omega_0^2}
{\cal A}(\omega_0)\sim \Lambda^{0}$ is logarithmically divergent. Two counterterms  bilinear in the field will be added in order to subtract the divergent parts.

\item  three-point graphs ($n=3$) are logarithmically divergent. A cubic counterterm will be added to subtract their divergent part.
\end{itemize}

Observe that, as long-range models do not exhibit a wave function renormalization,\footnote{In order to separate the divergent part of a graph we Taylor expand it at small momentum; but in a Taylor expansion only the integer powers of the momentum appear, hence the divergences are subtracted by counterterms for bilinear operators with an integer number of derivatives $\phi (-\partial^2)^{m} \phi$.} the coefficient of the $(p_0^2)^\zeta$ term in the two point function is not divergent. 

\paragraph{Renormalization.} In order to subtract the divergences we consider the renormalized action 
\begin{equation}
S_r[\phi]  = \int d \tau d^d x \Bigg\{ \frac{1}{2}\phi(\tau,x)
   \left[  \left(-\partial^2\right)^\z -  Z \; \partial_{\tau}^2\right]\phi(\tau,x) + \delta \kappa \;  \phi(\tau,x) + \frac{1}{2}  \delta m^{2\zeta} \, \phi(\tau,x)^2 + \frac{\imath }{3!} \lambda \, \phi(\tau,x)^3  \Bigg\} \; ,
\end{equation}
and the associated renormalized correlators for the renormalized field $\phi$. 

We will parameterize $\lambda = \mu^{\frac{\epsilon}{2} } g Z_g $, with $g$ the dimensionless renormalized coupling and $Z_g$ a multiplicative renormalization factor.
The  renormalization functions $Z =1 + \delta Z(g)$,
$Z_g = 1 + \delta Z_g( g)$ and $\delta m^{2\zeta}(g)$ and $\delta \kappa(g)$ are chosen so as to ensure that the renormalized correlators have no divergences.
The precise form of the counterterms depends on the renormalization scheme.
For example, we could fix them by imposing four BPHZ-like renormalization conditions, one for each class of divergent graphs. However, as usual for a massless theory, such conditions must be imposed at a nonvanishing subtraction momentum scale $\mu$. In this respect, we note that it is more practical to chose a configuration of external momenta in which the spatial momentum is set to zero and the external frequency acts as cutoff:\footnote{Due to the anisotropy of the problem, it is not obvious that this choice will prevent infrared divergence at all orders, but we conjecture that it is true.}
\be \label{eq:ren-cond}
\bar \Gamma_r^{(1)} = 0  \; , \qquad \bar \Gamma_r^{(2)}(0, 0) = 0
\; ,\qquad  \partial_{\omega_0^2} \bar \Gamma_r^{(2)}(\mu^\zeta, 0) =  1 
 \;,\qquad  \bar \Gamma_r^{(3)}(\mu^\zeta, 0 ; -(1+ c) \mu^\zeta,0)=\imath  \mu^{\frac{\epsilon}{2}}g \;.
\ee
The constant $c$ should be chosen so that the configuration is nonexceptional \cite{Kleinert:2001ax}, in our case meaning $c\neq 0$, and it is otherwise an arbitrary choice of the renormalization scheme, not affecting any observable quantity, such as the critical exponents.
However, at one loop we encounter no problem at $c=0$, and as calculations are easier in this case, we will make such exceptional choice in the following.

The first two conditions are automatically guaranteed in analytic regularization, because the propagator being massless, the amplitudes of the corresponding diagrams give pure power divergences (in the momentum cutoff), which are set to zero in analytic regularization. With other choices of regularization, we would instead fix $\delta m^{2\zeta}(g)$ and $\delta \kappa(g)$  so as to completely subtract the power divergent amplitudes. Either way we can safely ignore them from now on.

The remaining two conditions require in general subtractions of both divergent and finite terms.
In minimal subtraction we instead only subtract the divergent parts, hence $ \partial_{\omega_0^2} \bar \Gamma_r^{(2)}(\mu^\zeta, 0)$ and $ \bar \Gamma_r^{(3)}(\mu^\zeta, 0 ; -\mu^\zeta,0)$ will be more complicated functions of the renormalized coupling, but the structure of counterterms and renormalization group functions is simplified.
In the following we will use minimal subtraction.

We go to the weakly relevant case $\zeta = \frac{d+\epsilon}{5}$ and compute the amplitudes in analytic continuation in $\epsilon$. Besides setting the power divergences to zero, this has the effect of converting the logarithmic divergences in poles in $\frac{1}{\epsilon}$. Although $\eps$ is now to be kept finite, the same renormalization procedure as for the $\eps=0$ theory is to be carried out, as otherwise we would not be entitled to trust the small $\eps$ expansion.
In minimal subtraction the counterterms $\delta Z(g)$ and $\delta Z_g(g)$ are series in $1/\epsilon$ (that is they have no finite part) which are tuned so that the renormalized correlators have a well defined $\epsilon \to 0$ limit:\footnote{In practice this is enforced at a renormalization point, as above, but once the pole is taken out in such a way, the correlators $\Gamma_r^{(n)}(\omega,p)$ have a well defined $\epsilon \to 0$ limit for arbitrary external frequencies and momenta $\omega,p$.}
\begin{equation} \label{eq:counterterms}
    \delta Z(g) = \sum_{\ell\geq 1} \frac{\a_\ell(g)}{\eps^\ell}\;, \quad
    \delta Z_g(g) = \sum_{\ell\geq 1} \frac{\g_\ell(g)}{\eps^\ell}\;.
\end{equation}
The coefficients $\a_\ell(g)$ and $\g_\ell(g)$ are power series in $g$,  starting at order $g^{2\ell}$,\footnote{In order to show this, first one notices that order-$\ell$ poles appear first at $\ell$ loops, which in turn can for example be derived by the finiteness of beta function and anomalous dimension, as in \cite{ZinnJustin:2002ru}. Next, remembering that for an $n$-point graph with $P$ internal propagators, $V$ vertices and $\ell$ loops we have the relations $3V=2P+n$ and $\ell=P-V+1$, we find $V=2\ell+n-2$, from which, choosing $n=2$ and $3$ and remembering that vertex counterterms are multiplied by an extra factor of $g$, our statement follows.} with finite coefficients independent of $\eps$ or $\mu$.

The counterterms are treated as additional vertices  $\delta Z(g)$ and $\delta Z_g(g)$, and in the presence of such vertices the renormalized two and three point 1PI functions write at first nontrivial order in $g$ as:
\begin{equation}
\begin{split}
& \bar \Gamma_r^{(2)}(\omega_0,p_0)  =[ 1+\delta Z(g) ] \ \omega_0^2+ (p_0^2)^{\z}+\frac{\m^{\eps}}{2}  g^2 \ \mathcal{B}(\omega_0,p_0)\ ,  \crcr
&\bar \Gamma_r^{(3)}(\omega_1,p_1;\omega_2,p_2) = \imath \m^{\eps/2} g [ 1 +\delta Z_g(g) ]  - \imath  \m^{3 \eps/2}  g^3 \ \mathcal{T}(\omega_1,p_1;\omega_2,p_2) \; .
\end{split}
\end{equation}

Note that ${\cal B}$, ${\cal T}$, and all the other graph amplitudes arising in these expansions are computed with the bare propagator $  \omega^2 + (p^2)^\z$. The point is that the bare expansion generates divergences and these divergences are subtracted order by order by adding the counterterms: there is never any reason to compute amplitudes of graphs using the renormalized propagator $ [1+\delta Z(g)]\ \omega^2 + (p^2)^\z$. Indeed, using the renormalized propagator corresponds to resummations of infinite families of bare graphs with arbitrary insertions of the counterterm vertex $\delta Z(g)$ on all the edges. 

For the bubble and the triangle graphs we have the following singular behaviour:
\be \label{eq:bt}
\partial_{\omega_0^2}{\cal B}(\mu^\z,0) = -\mu^{ -\eps}  \frac{  b}{\epsilon}  + \text{finite}
 \;,\qquad {\cal T}(\mu^\z,0;-\mu^\z,0) =
 \mu^{-\eps} \frac{ t }{\epsilon}  + \text{finite} \;.
\ee
where $b$ and $t$ are given in Appendix \ref{app:Integrals}, hence the minimal subtraction counterterms that ensure that the one-loop divergences are subtracted are:
\begin{equation}
    \delta Z(g) = \frac{1}{2}  g^2  \; \frac{b}{\epsilon}  \;,\qquad
     \delta Z_g(g) = g^2  \; \frac{t}{\epsilon} \;.
\end{equation}

\paragraph{Renormalized theory from the bare theory.} Up to terms which we can ignore, the renormalized theory is described by the action:
\begin{equation} \label{eq:S_renorm}
    S[\phi] = \int d \tau d^d x \Bigg\{ \frac{1}{2}\phi(\tau,x)
   \left[  \left(-\partial^2\right)^\z - Z \partial_{\tau}^2\right]\phi(\tau,x)+\imath \frac{\lambda}{3!} \phi(\tau,x)^3  \Bigg\} \; ,
\end{equation}
and maps onto the bare one by a field redefinition and change of couplings:\footnote{\label{foot:alpha-family}In fact, because of the scale invariance of the free theory, a one parameter family of mappings onto different bare theories exist, involving a rescaling also of $\tau$ and $x$: 
\[
  \phi(\tau,x) = Z^{ - \frac{1}{4} + \alpha \Delta_\phi} \phi^{(1)}_b(Z^{-\frac{1}{2} + \z \alpha}\tau,Z^{\alpha}x  ) 
  \; , \;\;  \lambda Z^{-\frac{1}{4} - (d+\zeta-3\Delta_\phi) \alpha} = \lambda^{(\alpha)}_b 
  \qquad S[\phi;\lambda,Z] = S_b[\phi^{(\alpha)}_b, \lambda^{(\alpha)}_b] \; .
\]
We can interpret this as a family of renormalization schemes, which is more evident if we keep fixed the bare coupling.
Remembering that $d+\zeta-3\Delta_\phi=\eps/2$, only the $\a=0$ scheme corresponds to a true minimal subtraction, in the sense that the relation between $\lambda$ and $\lambda_b$ does not involve finite redefinitions of the coupling,
$\lambda_b = \m^{\eps/2} g \left(1+ \sum_{n\geq 1} \frac{\tilde \g_n(g)}{\eps^n}  \right)$.
For $\a\neq 0$, the expansion in powers of $g$ of the factor $Z^{-\a \eps/2}$ leads to positive powers of $\eps$ that introduce in the above relation new terms that are finite or vanishing in the $\eps\to 0$ limit.
}
\begin{equation}
\begin{split}
  \phi(\tau,x) = Z^{ - \frac{1}{4} } \phi_b(Z^{-\frac{1}{2}}\tau,x  ) 
  \; , \;\;  \lambda Z^{-\frac{1}{4}} = \lambda_b 
  \qquad &\Rightarrow \qquad S[\phi;\lambda,Z] = S_b[\phi_b, \lambda_b] \; .
\end{split}
\end{equation}
The renormalized correlators are related to the bare  correlators by:
\begin{equation} \label{eq:F_r-b}
\begin{split}
F_r^{(n)}(\tau,  x | \lambda ,Z ) 
 = \frac{\int [d\phi] e^{-S(\phi) } \; \phi( \tau_1,x_1) \dots
 \phi( \tau_{n},x_{n} )  }{ \int [d\phi] e^{-S(\phi) } } 
&  = Z^{-\frac{n}{4} } \; F^{(n)}_b( Z^{-\frac{1}{2}} \tau, x | \lambda Z^{-\frac{1}{4}} ) 
\; ,
 \end{split}
\end{equation}
and we note that, contrary to the more familiar isotropic case, one needs to rescale the time argument with an appropriate power of $Z$.

One computes the renormalized amputated 1PI correlators of the theory using the renormalized action $S[\phi]$, or by a change of variables in terms of the bare one:
\begin{equation}
e^{-\Gamma_r[\Phi]} =  \int_{\rm 1PI} d\varphi \; e^{-S[\Phi +\varphi;\lambda,Z]} 
= \int_{\rm 1PI} d\varphi \; e^{-S_b[\Phi_b +\varphi; \lambda Z^{-\frac{1}{4} } ]}  \;,\qquad
  \Phi(\tau,x) = Z^{-\frac{1}{4}} \Phi_b( Z^{-\frac{1}{2}} \tau,x ) \;,
\end{equation}
that is the relation between the bare and renormalized effective actions reproduces the relation between the classical actions, $\Gamma_r[\Phi|\lambda,Z]= \Gamma_b[ \Phi_b | \lambda Z^{-\frac{1}{4}} ]|_{\Phi_b(\tau,x) = Z^{\frac{1}{4} } \Phi( Z^{1/2} \tau ,x) }$. This implies that the amputated correlators transform like:\footnote{We use 
$ \int d\tau dx \; \Phi(\tau,x)^n\Gamma_r^{(n)} (\tau,x| \lambda,Z)  = \int d\tau' dx' \;
 \left(Z^{\frac{1}{4} } \Phi( Z^{1/2} \tau' ,x') \right)^n \Gamma_b^{(n)}( \tau',x' | \lambda Z^{-\frac{1}{4}} )$ and change variable.
}
\begin{equation}
\Gamma_r^{(n)} (\tau,x|\lambda,Z) 
 = Z^{- \frac{n}{4}} \Gamma_b^{(n)} ( Z^{-\frac{1}{2}}\tau, x  | \lambda Z^{-\frac{1}{4}} )
\; ,
\end{equation}
or in momentum space:
\begin{align}
   \Gamma_r^{(n)}(\omega,p|\lambda,Z) = Z^{ \frac{n}{4}} \Gamma_b^{(n)} ( Z^{\frac{1}{2}}\omega, p  | \lambda Z^{-\frac{1}{4}} ) \;, \qquad 
\bar \Gamma_r^{(n)}(\omega,p|\lambda,Z) = Z^{ \frac{n}{4} - \frac{1}{2}} \bar \Gamma_b^{(n)} ( Z^{\frac{1}{2}}\omega, p  | \lambda Z^{-\frac{1}{4}} )  \; .
\end{align}

As a sanity check, we can compute the amplitudes of graphs reorganizing the renormalized theory, that is including the full coupling $Z$ in the propagator. The amplitudes of amputated graphs contributing to the renormalized amputated correlators write as: 
\begin{equation}
\begin{split}
{\cal \tilde A}(\omega_E,q_E) & = \int \prod_{L} d\omega_L d^dp_L \; \prod_{e} \frac{1}{ Z  \, \omega_e^2 + (q_e^2)^\z } 
 \xlongequal{\omega_L = \omega_L' Z^{-1/2} } \left(    Z^{-1/2}   \right)^{P-V+1} {\cal A} (Z^{1/2} \omega_E,q_E)\crcr
 & = Z^{-V/4 +n/4 -1/2} {\cal A} (Z^{1/2} \omega_E,q_E)
\; , \qquad {\cal A}(\omega_E,q_E) = \int \prod_{L} d\omega_L d^dp_L \; \prod_{e} \frac{1}{ \omega_e^2 + (p_e^2)^\z } \;,
\end{split}
\end{equation}
with ${\cal A}(\omega_E,q_E)$ being the bare amplitude with $Z=1$. Since each such amplitude is multiplied by $\lambda^V$, we recover the above relation between bare and renormalized proper vertices:
\begin{equation} \label{eq:properVertices_b-r}
\bar \Gamma_r^{(n)}(\omega_E,q_E|\lambda,Z) = Z^{\frac{ n}{4} - \frac{1}{2} } \; \bar \Gamma_b^{(n)}(Z^{1/2}\omega_E,q_E|\lambda Z^{ -1/4 }) \; .
\end{equation}

We note that, contrary to the more familiar isotropic case, the relation between the bare and renromalized versions of the general correlators and of the amputated one particle irreducible ones is the same. 

\paragraph{The renormalization group flow.} The renormalized correlators $ F_r^{(n)} (\tau,x|\mu^{\frac{\epsilon}{2}} gZ_g ,Z)  $ are functions only of $g$ and $\mu$ (and of course $\tau$ and $x$) and are free of divergences. 
To emphasize this fact, and to follow common practice, we will write the renormalized  correlators as $ F_r^{(n)} (\tau,x|g,\mu)  $, with a slight abuse of notation. 
In terms of such new notation, we rewrite equation \eqref{eq:F_r-b} as:
\begin{align} \label{eq:br-F_n}
    F_r^{(n)} (\tau,x|g,\mu) 
 = Z^{- \frac{n}{4}} F_b^{(n)} ( Z^{-\frac{1}{2}}\tau, x  | \mu^{\frac{\epsilon}{2}} gZ_g Z^{-\frac{1}{4}} ) \;,
 \end{align}
where at fixed $g$ and $\eps>0$ both the renormalized and the bare correlators depend on $\mu$ via the explicit combination on the right-hand side of the equation. At $\eps=0$, due to the poles in $1/\epsilon$ of the bare correlators and renormalization functions, a logarithmic dependence on $\mu$ survives. 

The renormalization group is designed to capture the $\mu$-dependence of renormalized correlators for a fixed bare theory, and it does so with the introduction of beta function and anomalous dimension.
The beta function  $\beta = \mu \frac{dg}{d\mu} $  is obtained by 
tuning $g$ with $\mu$ so that the bare coupling stays fixed: 
\begin{align} 
  \mu \frac{d}{d\mu} \left[ \mu^{\frac{\epsilon}{2}} gZ_g Z^{-\frac{1}{4}}\right] =0 \qquad \Rightarrow \qquad 
 \beta(g) = -\frac{\frac{\epsilon}{2}g}{ 1 + g\frac{Z'_g}{Z_g} -\frac{1}{4} g \frac{Z'}{Z}  }  \;,
 \end{align}
 where a prime denotes a derivative with respect to $g$. 
Using the expansion \eqref{eq:counterterms} in powers of $1/\eps$ for the counterterms, we find
\begin{equation}
    \b(g) = -\frac{\epsilon}{2} g +\frac{g^2}{2} \left(\g_1'(g)-\frac{\a_1'(g)}{4}\right) \; ,
\end{equation}
and all the contributions from $\a_n$ and $\g_n$ with $n>1$, as well as higher powers of the $n=1$ terms, must cancel for consistency.
 At one loop we have $\a_1(g) =  \frac{1}{2} g^2 \;  b$ and $ \g_1(g)  = g^2  \; t$ hence we get:\footnote{In the $\a$-dependent scheme of footnote \ref{foot:alpha-family} we would obtain $\beta= -\frac{\epsilon}{2} g + g^3 ( t -b (1+2\epsilon\a)/8)$. The scheme dependence of the one-loop term is not surprising as it is scheme-independent only in the marginal case, $\eps=0$. The critical exponents are instead $\a$-independent, as they should.}
\begin{equation}
    \beta(g)  
     = -\frac{\epsilon}{2} g + g^3 ( t -\frac{b}{8}) +O(g^5) \;.
\end{equation}
Such beta function has a zero, i.e.\ a fixed point, at (note that $t=3b$ according to Appendix~\ref{app:Integrals}):
\be
    g_\star =  \sqrt{ \frac{ \epsilon }{  2 t-\frac{b }{4} } }  =  2^{\frac{d+5}{2}} \pi ^{d/4} \sqrt{\frac{\epsilon \, \Gamma \left(\frac{d}{2}\right)}{23} }  \;,
\ee
which is infrared attractive, as the corresponding correction-to-scaling exponent is positive: 
\be
\partial_g \beta(g) |_{g=g^\star}=\eps \; .
\ee

For the anomalous  dimension $\eta = \mu\tfrac{d}{d\mu}\ln Z = \beta Z'/Z$, at one loop we find $\eta = - g^2 b$, which at the fixed point becomes:
\be
\eta^{\star} = -\frac{\epsilon b}{  2 t-\frac{b}{4} } = -\frac{4 \epsilon }{23} \; .  
\ee

 The Callan-Symanzik equation encapsulates the change of the  renormalized correlators with the renormalization scale. It is obtained by taking into account that the bare correlators do not depend on the renormalization scale, $\mu \frac{d}{d\mu} F_b^{(n)} (\tau, x  | \mu^{\frac{\epsilon}{2}} gZ_g Z^{-\frac{1}{4}} ) =0 $, and using \eqref{eq:br-F_n}. 
 Due to the time rescaling between the renormalized and the bare correlator, one gets an additional term with respect to the usual Callan-Symanzik equation:
\begin{align}
\mu \frac{d}{d\mu} F_r^{(n)}  = 
-\frac{n}{4} \eta \; F_r^{(n)} -\frac{1}{2}\eta 
D_\tau F_r^{(n)} \quad
\Rightarrow \quad \left[ \mu\partial_\mu + 
  \beta \partial_g + \frac{n}{4}\eta + \frac{\eta}{2}D_\tau
\right] F_r^{(n)} = 0 
\end{align}
where $D_\tau = \sum_i \tau_i \partial_{\tau_i}$ is the time dilatation operator. We note that, contrary to the isotropic case, one gets the exact same form of the Callan-Symanzik  for the one particle irreducible amputated correlator. 

\paragraph{Lifshitz scaling at fixed point.}
In order to understand the behaviour of the renormalized correlators 
(amputated or not) under rescaling we combine 
Eq.~\eqref{eq:br-F_n} and Eq.~\eqref{ScaleTrasfGn} to conclude 
\begin{equation}
\begin{split} \label{eq:Fn-dimAnalysis}
 F_r^{(n)} (\tau,x|g,\mu) 
& =  l^{n \Delta_\phi}  F_r^{(n)} (l^\z\tau,l x| g,l^{-1}\mu)  \;,
\end{split}
\end{equation}
which again is just a statement about engineering dimensions.
In order to deduce scaling dimensions, we 
act on both sides with the derivative $l\frac{d}{dl}$, evaluated at $l=0$, and we combine it with the Callan-Symanzik equation, obtaining:\footnote{
Taking into account that the amputated correlator changes under rescaling according to Eq.~\eqref{eq:gammabareresc}, an equation similar to Eq.~\eqref{eq:dilatF} with $ \Delta_\phi$ replaced by $d+\zeta - \Delta_\phi $ holds in that case. The two equations are consistent as $d+\zeta -\Delta_\phi + \frac{\eta}{4}=
d + (\zeta + \frac{\eta}{2}) - (\Delta_\phi + \frac{\eta}{4} ) $.}
\be\label{eq:dilatF}
\begin{split}
 \bigg[ n \Delta_\phi + \zeta D_\tau + D_x- \mu \partial_\mu
      \bigg] F_r^{(n)} = 0  \quad
\Rightarrow \quad \bigg[ n(\Delta_\phi + \frac{\eta}{4}) + 
( \zeta + \frac{\eta}{2} ) D_\tau + D_x + \beta\partial_g  \bigg] F_r^{(n)} = 0 \;.
\end{split}
\ee
At the fixed point, the correlators are eigenfunctions of the anisotropic scaling operator $z D_\tau + D_x$, with the following dynamic exponent and scaling dimension:
\begin{equation}
\begin{split}
    &z = \z + \frac{\eta_\star}{2} = \f{d}{5}+ \frac{13 \eps}{115}  \;, \qquad  \\
    &\Delta_\star = \Delta_\phi + \frac{\eta_\star}{4} = \frac{2d}{5} -  \frac{33\eps}{230} \; .
\end{split}
\end{equation}
That is, they have the following Lifshitz scale invariance
\begin{equation}  \label{eq:Fn-scaling}
F_r^{(n)} (\tau,x|g_\star, \mu)
= l^{n\Delta_\star}  F_r^{(n)} ( l^{z} \tau, l x|g_\star, \mu) \;.
\end{equation}
For example, at $n=2$  the solution of the fixed point scaling equation is:
\begin{equation} \label{eq:2pt-scaling}
    F_r^{(2)} (\tau,x|g_\star, \mu)  = |x|^{-2\Delta_\star} \mu^{-\eta_\star/2} f\left( \f{\t^2}{\mu^{\eta_\star}|x|^{2z}} \right) \ ,
\end{equation}
where the function $f(u)$ is regular in zero and vanishes at infinity as $u^{-\Delta_\star/z}$, and otherwise it is completely unconstrained. The explicit factors of $\m$ in \eqref{eq:2pt-scaling} are there to make it compatible with canonical dimensional analysis, summarized in \eqref{eq:Fn-dimAnalysis}, but play no role for the more interesting scaling given by \eqref{eq:Fn-scaling}.

In order to compare to existing literature on Lifshitz field theories, it is instructive to repeat the above analysis for the proper vertices in momentum space. Recalling that the renormalized vertices are written in terms of the bare ones as
\begin{equation}
    \bar \Gamma_r^{(n)}(\omega,p|g,\mu) = Z^{ \frac{n}{4} - \frac{1}{2}} \bar \Gamma_b^{(n)} ( Z^{\frac{1}{2}}\omega, p  | \mu^{\frac{\epsilon}{2}} gZ_g Z^{-\frac{1}{4}}) \;,
\end{equation}
we deduce the Callan-Symanzik equation:
\begin{equation}
     \left[ \mu\partial_\mu + 
  \beta \partial_g + \Big(\f12-\frac{n}{4}\Big)\eta - \frac{\eta}{2} \omega \p_\omega \right] \bar \Gamma_r^{(n)}(\omega,p|g,\mu)   = 0 \;.
\end{equation}
Dimensional analysis gives instead
\begin{equation}
    \bar \Gamma_r^{(n)} (\omega,p |g,\mu) = l^{\zeta+d} l^{ - n \Delta_\phi} \bar \Gamma_r^{(n)}(l^{-\z} \omega, l^{-1} p | g,l^{-1}\mu ) \;,
\end{equation}
which combined with the Callan-Symanzik equation leads to
\be\label{eq:dilatGamma}
\begin{split}
 &\bigg[  (\z+d-n\Delta_\phi) - \zeta D_\omega - D_p- \mu \partial_\mu  \bigg] \bar \Gamma_r^{(n)}(\omega,p|g,\mu) = 0 \\
 & \quad \Rightarrow \quad \bigg[  \z+d +\f{\eta}{2} -n(\Delta_\phi+ \frac{\eta}{4}) - ( \zeta + \frac{\eta}{2} ) D_\omega - D_p + \beta\partial_g   \bigg] \bar \Gamma_r^{(n)}(\omega,p|g,\mu) = 0 \;.
\end{split}
\ee
For $n=2$, at the fixed point, this becomes:
\begin{equation}
    \bigg[  2\z  - z D_\omega - D_p   \bigg] \bar \Gamma_r^{(2)}(\omega,p|g_\star,\mu) = 0 \;,
\end{equation}
whose solution is
\begin{equation}
    \bar \Gamma_r^{(2)} (\omega,p|g_\star,\mu)  = |p|^{2\z}  f\left( \f{\omega^2}{\mu^{-\eta_\star}|p|^{2z}} \right) \;,
\end{equation}
where the function $f(u)$ is regular in zero and blows up at infinity as $u^{\z/z}$.
Therefore, we have
\begin{equation}
    \begin{split}
        &\bar \Gamma_r^{(2)} (0,p|g_\star,\mu)  \sim |p|^{2\z} \equiv |p|^{2\z-\eta_{2\z}}\;,\\
        &\bar \Gamma_r^{(2)} (\omega,0|g_\star,\mu)  \sim \omega^{2\z/z} \equiv \omega^{2-\eta_2}\;,
    \end{split}
\end{equation}
where we introduced the standard definition of correlation exponents $\eta_2$ and $\eta_{2\z}$, for which we read off:
\begin{equation}
    \eta_2= 2-\f{2\z}{z} =\f{\eta_\star}{\z+\eta_\star/2} \;, \qquad \eta_{2\z} = 0\; .
\end{equation}
Lastly, we obtain the following expression for the dynamic exponent:
\begin{equation}
    z = \f{2\z-\eta_{2\z}}{2-\eta_2} \;,
\end{equation}
which holds also for integer values of $\z>1$, but with a nonvanishing $\eta_{2\z}$. 

As a final remark, we notice that for integer $\z$, besides the additional counterterms that need to be introduced, a different renormalization scheme is typically employed \cite{Diehl:2000sv,SHPOT2001340}, in which the bare and renormalized fields are simply related by $\phi(\tau,x) = Z_{\phi}^{ - \frac{1}{2} } \phi_b(\tau,x  ) $, with no rescaling of $\t$. However, a further rescaling is later performed on $\phi$ and $x$ in order to remove the redundant coupling associated to the spatial higher-derivative operator ($\sigma$ in the notation of \cite{Diehl:2000sv,SHPOT2001340}). It can be shown that the net effect of such operations corresponds to the choice $\a=1/(2\z)$ in the $\a$-dependent scheme of footnote \ref{foot:alpha-family}. Therefore, universal quantities are not affected by such choice.

\section{The quartic model} 
\label{sec:quartic}

The results generalize mutatis mutandis to a model with quartic potential:

\begin{equation}
    S_b[\phi_b] = \int d \tau d^d x \Bigg\{ \frac{1}{2}\phi_b(\tau,x)
   \left[  \left(-\partial^2\right)^\z - \partial_{\tau}^2\right]\phi_b(\tau,x)+ \frac{\lambda_b}{4!} \phi_b(\tau,x)^4  \Bigg\} \; ,
\end{equation}
 While the field dimension is still $\Delta_\phi = \frac{d-\zeta}{2}$, the coupling has dimension $d+\zeta -4\Delta_\phi$. 
 Therefore, the quartic interaction is marginal for $\zeta =\frac{d}{3}$ and weakly relevant for $\zeta =\frac{d+\epsilon}{3} $, corresponding to $d+\zeta -4\Delta_\phi = \epsilon$.
Therefore, in order to have an interesting renormalization group flow, and keep $\z<1$, we must stick to $d<3$.

We can rephrase the statement about marginality from the point of view of power counting, which leads like before to the superficial divergence $\Lambda^{(d+\zeta)(P-V+1) - 2\zeta P}$. However, this time we have $2P=4V-n$, and thus the superficial divergence is $  \Lambda^{d+\zeta - \frac{n}{2} (d-\zeta) + V (d -3 \zeta ) }$, 
which for $\zeta = \frac{d+\epsilon}{3}$ becomes $\Lambda^{\frac{d}{3} (4 - n)-\eps (V -\f{2+n}{6}) }$. Therefore, for $\eps=0$ the theory is renormalizable, with logarithmically-divergent four-point point graphs, while for $\eps>0$ it is super-renormalizable, with finite four-point graphs.

The bare full correlators and the bare 1PI correlators behave under rescaling as:
 \begin{align}
 &F_b^{(n)}(\tau,x | \lambda_b ) =  l^{n \Delta_\phi } F_b^{(n)}(l^\zeta \tau, l x | l^{- \epsilon }\lambda_b )   \;, \;\;
\bigg[ n\Delta_\phi  - \epsilon  \lambda_b \partial_{\lambda_b} + \zeta D_\tau + D_x \bigg] F_b^{(n)}(\tau,x | \lambda_b ) = 0 \;, \\
&\Gamma_b^{(n)}(\tau,x | \lambda_b ) =  l^{n (\zeta + d - \Delta_\phi)} \Gamma_b^{(n)}(l^\zeta \tau, l x | l^{- \epsilon }\lambda_b )   \;, 
\;\;
\bigg[ n(\zeta + d -\Delta_\phi)  - \epsilon \lambda_b \partial_{\lambda_b} + \zeta D_\tau + D_x \bigg] \Gamma_b^{(n)}(\tau,x | \lambda_b ) = 0 \;. \nonumber
\end{align}

\begin{figure}
    \centering
    \includegraphics[width=0.5\textwidth]{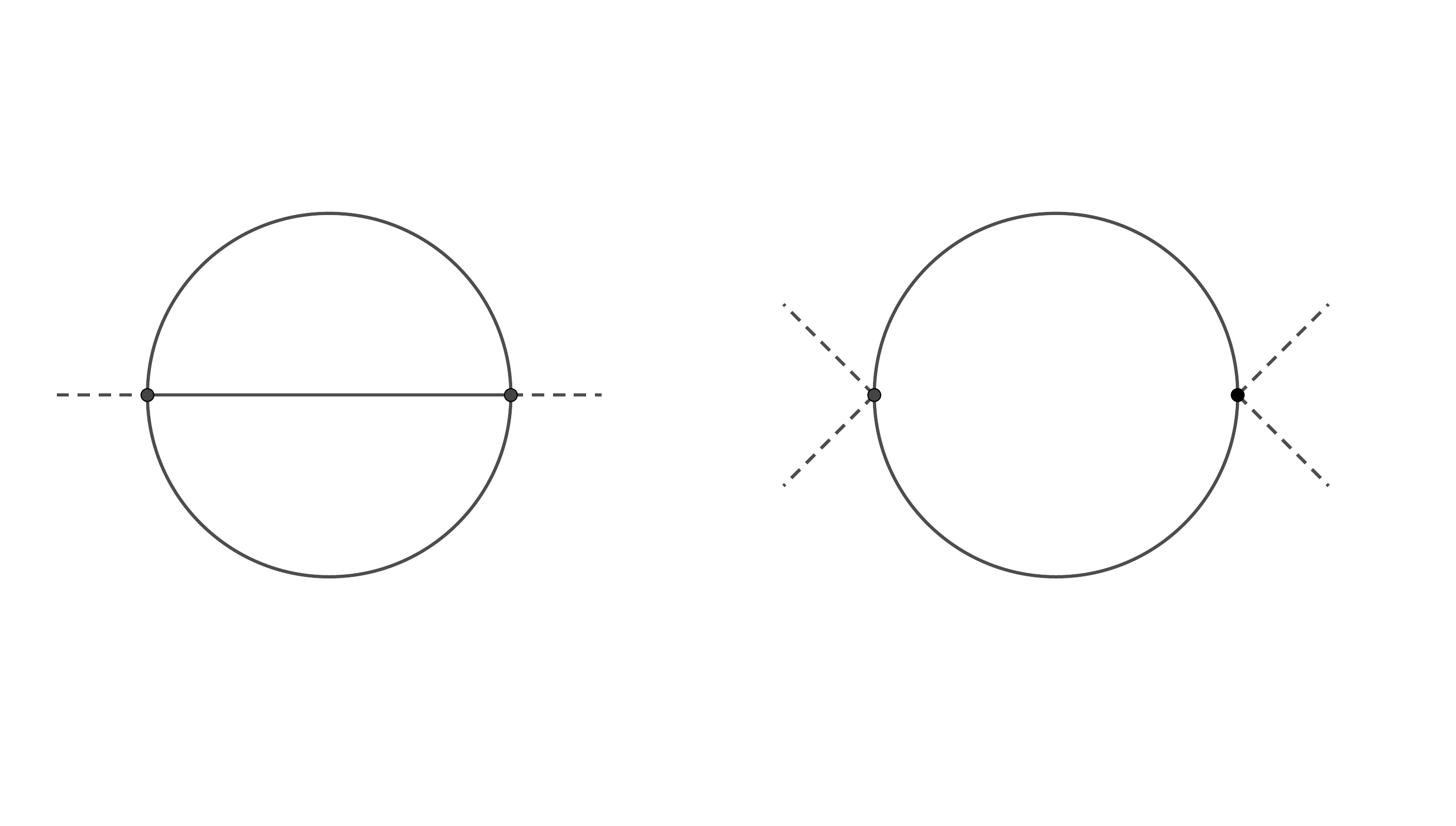}
    \caption{Left: melon diagram $\mathcal{M}(\omega_0,p_0)$. Right: bubble diagram $\mathcal{B}(\omega_1+\omega_2,p_1+p_2)$.}
    \label{fig:elonBubble}
\end{figure}

Due to $\mathbb{Z}_2$ invariance of the model, only the correlators with even $n$ are non-vanishing.
For the $n=2$ and $4$ correlators in momentum space, with the global momentum and frequency conservation factored out, we have at one loop:
\begin{equation}
\begin{split}
& \bar \Gamma_b^{(2)}(\omega_0,p_0)  =\omega_0^2+ (p_0^2)^{\z}
-\frac{1}{6}\lambda_b^2 \ \mathcal{M}(\omega_0,p_0)\ , \\
& \bar \Gamma_b^{(4)}(\omega_1,p_1;\omega_2,p_2;\omega_3,p_3) = \lambda_b - \frac{1}{2} \lambda_b^2 \ \Big( \mathcal{B}(\omega_1+\omega_2,p_1+p_2) \\
&\hspace{5cm} +\mathcal{B}(\omega_1+\omega_3,p_1+p_3)+\mathcal{B}(\omega_2+\omega_3,p_2+p_3)\Big) \ ,
\end{split}
\end{equation}
where $\mathcal{M}$ and $\mathcal{B}$ are respectively the melon and the bubble diagrams depicted in Fig.~\ref{fig:elonBubble}
with amplitudes:
\begin{align}
& \mathcal{M}(\omega_0,p_0) = \int \frac{d^dq_1 \; d\omega_1}{(2\pi)^{d+1} }  \int \frac{d^dq_2 \; d\omega_2}{(2\pi)^{d+1} }  
\;\frac{1}{ \omega_1^2 + (q_1^2)^\zeta } \; 
\;\frac{1}{ \omega_2^2 + (q_2^2)^\zeta } \; 
\frac{1}{ (\omega_0+\omega_1+\omega_2)^2 + 
[(p_0 + q_1+q_2)^2 ]^\zeta } 
\; ,
\crcr 
&  \mathcal{B}(\omega_0,p_0) = \int \frac{d^dq \; d\omega}{(2\pi)^{d+1} }  
\;\frac{1}{ \omega^2 + (q^2)^\zeta } \; \frac{1}{ (\omega_0+\omega)^2 + 
[(p_0 + q)^2 ]^\zeta } \; .
\end{align}
As renormalization point we will take $(\omega_1,p_1)=(\omega_2,p_2)=(\omega_3,p_3)=(\mu^\z/2,0)$ and  $(\omega_4,p_4)=(-\f32 \mu^\z,0)$, so that $\omega_1+\omega_2=\omega_1+\omega_3=\omega_2+\omega_3=\m^{\z}$ and $\sum_i \omega_i = 0$.

Notice that the bubble integral is exactly as the one we encountered for the two-point function of the cubic model, but in the present case it is to be evaluated in a different range of $\z$, and it is now associated to a four-point function. The net effect is that in the quartic model we are not interested in $\partial_{\omega_0^2}{\cal B}(\mu^\z,0)$, which in this case converges for $\eps\to 0$, but rather in ${\cal B}(\mu^\z,0)$ itself, which has a $1/\eps$ pole.

The renormalized action is:
\be
S_r[\phi]  = \int d \tau d^d x \Bigg\{ \frac{1}{2}\phi(\tau,x)
   \left[  \left(-\partial^2\right)^\z -  Z \; \partial_{\tau}^2\right]\phi(\tau,x) + \frac{1}{2}  \delta m^{2\zeta} \, \phi(\tau,x)^2 + \frac{1}{4!} \lambda \, \phi(\tau,x)^4  \Bigg\} \; ,
\ee
with $\lambda = \mu^\epsilon gZ_g$ and counterterms 
$\delta Z(g)=Z-1$, $\delta m^{2\zeta} $ and $\delta Z_g(g)=Z_g-1$. 
We will again employ analytic regularization and ignore the mass counterterm. For the other two counterterms we have, in the minimal subtraction scheme, an expression like \eqref{eq:counterterms}, where however the coefficients $\a_\ell(g)$ and $\g_\ell(g)$ are now starting at order $g^{\ell+1}$ and $g^\ell$, respectively.\footnote{In this case order-$\ell$ poles appear first at $\ell+1$ loops, for $\p_{\omega^2}\bar \Gamma_b^{(2)}$, and  at $\ell$ loops, for $\bar \Gamma_b^{(4)}$. Moreover, the topological relation is now $4V=2P+n$.}

The renormalized 1PI two and four-point functions write at first nontrivial order in $g$:
\begin{equation}
\begin{split}
& \bar \Gamma_r^{(2)}(\omega_0,p_0)  =[ 1+\delta Z(g) ] \ \omega_0^2+ (p_0^2)^{\z} - \frac{1}{6}\m^{2\eps}  g^2 \ \mathcal{M}(\omega_0,p_0)\ ,  \crcr
&\bar \Gamma_r^{(4)}(\omega_1,p_1;\omega_2,p_2;\omega_3,p_3) =\m^{\eps} g [ 1 +\delta Z_g(g) ]  - \frac{1}{2} \m^{2 \eps}  g^2 \ \Big( \mathcal{B}(\omega_1+\omega_2,p_1+p_2) \\
&\hspace{6cm} +\mathcal{B}(\omega_1+\omega_3,p_1+p_3)+\mathcal{B}(\omega_2+\omega_3,p_2+p_3)\Big) \ ,
\end{split}
\end{equation}
with singular behaviour:
\be \label{eq:BM-phi4}
\partial_{\omega_0^2}{\cal M}(\mu^\z,0) = -\mu^{ -2\eps}  \frac{  M}{\epsilon}  + \text{finite}
 \;,\qquad {\cal B}(\mu^\z,0) =
 \mu^{-\eps} \frac{ B }{\epsilon}  + \text{finite} \;,
\ee
with $B$ and $M$ given in appendix \ref{app:Integrals}, hence 
the minimal subtraction counterterms that ensure that the divergences are subtracted are:
\begin{equation}
    \delta Z(g) = - \frac{1}{6} \, g^2  \; \frac{M}{\epsilon}  \;,\qquad
     \delta Z_g(g) = \frac{3}{2} \, g  \; \frac{B}{\epsilon} \;.
\end{equation}

The renormalized theory maps onto the bare one by the following field redefinition and change of couplings:\begin{equation}
\begin{split}
  \phi(\tau,x) = Z^{ - \frac{1}{4} } \phi_b(Z^{-\frac{1}{2}}\tau,x  ) 
  \; , \qquad  \lambda Z^{-\frac{1}{2}} = \lambda_b   \;.
\end{split}
\end{equation}
The structure of the beta function is slightly altered with respect to the cubic case:
\begin{align} 
  \mu \frac{d}{d\mu} \left[ \mu^{\epsilon } gZ_g Z^{-\frac{1}{2}}\right] =0 \qquad \Rightarrow \qquad 
 \beta = -\frac{\epsilon g}{ 1 + g\frac{Z'_g}{Z_g} -\frac{1}{2} g \frac{Z'}{Z}  } \;, \qquad \eta = \beta \frac{Z'}{Z} \;,
 \end{align}
 where again a prime denotes a derivative with respect to $g$. 
Using the expansion \eqref{eq:counterterms} in powers of $1/\eps$ for the counterterms, we find
\begin{equation}
    \b(g) = -\frac{\epsilon}{2} g +\frac{g^2}{2} \left(\g_1'(g)-\frac{\a_1'(g)}{2}\right) \; ,
\end{equation}
and all the contributions from $\a_n$ and $\g_n$ with $n>1$, as well as higher powers of the $n=1$ terms, must cancel for consistency.
At leading order we have $\a_1(g)= -  \frac{1}{6} g^2 \, M$ and $ \g_1(g)  = \frac{3}{2} \, g  \, B$, hence we get:\footnote{The melon contribution, corresponding to a two-loops diagram, does not enter the beta function in our leading-order approximation: it would contribute at order $g^3$, together with contributions from two-loops four-point diagrams. It is common to consider a strict one-loop approximation, so that the anomalous dimension is vanishing; however, because of such vanishing, it is also consistent to consider $\beta(g)$ and $\eta(g)$ at the same order in $g$, as we do here, rather than at the same loop order.}
\begin{equation}
    \beta(g) =
    -\epsilon g + \frac{3}{2} g^2 B +O(g^3)\;,\qquad
    \eta(g) = \frac{1}{3} \, g^2 M +O(g^3) \; ,
\end{equation}
with fixed point $ g_\star = \frac{2 \epsilon}{ 3 B}$, and anomalous dimension $\eta_\star = \frac{4 \epsilon^2 M}{27 B^2} $. 

The renormalized correlators write in terms of bare ones as:
\begin{align}
    F_r^{(n)} (\tau,x|g,\mu) 
 = Z^{- \frac{n}{4}} F_b^{(n)} ( Z^{-\frac{1}{2}}\tau, x  | \mu^{\epsilon} gZ_g Z^{-\frac{1}{2}} ) \;,
 \end{align}
and hence the Callan-Symanzik equation is unchanged.
In particular, we find again
\begin{equation}
    \bigg[ n(\Delta_\phi + \frac{\eta}{4}) + 
( \zeta + \frac{\eta}{2} ) D_\tau + D_x + \beta\partial_g  \bigg] F_r^{(n)} = 0 \; .
\end{equation}
Recalling that $\Delta_\phi = \frac{d-\zeta}{2}$, we conclude as in the cubic case that at the fixed point the correlators are eigenfucntions of the anisotropic scaling operator $z D_\tau + D_x$,  with the following dynamic exponent and scaling dimension:
\begin{equation}
\begin{split}
    &z = \z + \frac{\eta_\star}{2} = \f{d+\eps}{3}+\frac{2 \epsilon^2 M}{27 B^2} \;, \qquad  \\
    &\Delta_\star = \Delta_\phi + \frac{\eta_\star}{4} = \frac{2d-\eps}{6} + \frac{ \epsilon^2 M}{27 B^2} \; ,
\end{split}
\end{equation}
with the constants $B$ and $M$ given in \eqref{eq:B} and \eqref{eq:M_2}, respectively.

\section*{Acknowledgments}
R.~G.~and D.~L.~are supported by the European Research Council (ERC) under the European Union's Horizon 2020 research and innovation program (grant agreement No 818066) and by the Deutsche
Forschungsgemeinschaft (DFG) under Germany's Excellence Strategy EXC--
2181/1 -- 390900948 (the Heidelberg STRUCTURES Cluster of Excellence).

\newpage

\appendix

\section{Integrals} \label{app:Integrals}

\subsection{The bubble integral, $\mathcal{B}$}
The bubble integral writes:
\begin{equation}
\begin{split}
\mathcal{B}(\omega_0,p_0) &=
 \int \frac{d^dq \; d\omega}{(2\pi)^{d+1} }  
\;\frac{1}{\omega^2 + (q^2)^\zeta } \; \frac{1}{ (\omega_0+\omega)^2 + 
[(p_0 + q)^2 ]^\zeta }  \; .
\end{split}
\end{equation}

The integral is in general quite challenging, but when $p_0=0$ one can integrate $\omega$ by deforming the contour in the complex plane and using the residue theorem, and then use the spherical symmetry of the spacial momentum integral to obtain:
\begin{equation}
    \mathcal{B}(\omega_0,0)=  \int \frac{d^d q}{(2\pi)^d}\frac{(q^2)^{-\zeta/2}}{4(q^2)^{\zeta}+ \omega_0^2}  = -\frac{2^{1-\frac{d}{\zeta}}
     \pi \omega_0^{ \frac{d}{\zeta}-3}
    }{(4\pi)^{\frac{d}{2} } \Gamma(\frac{d}{2} )  \zeta \cos(\frac{\pi d}{2\z})} \; .
\end{equation}
In the evaluation of the spatial integral we have assumed  $\z\in(d/3,d)$, where it is convergent. Analytic regularization amounts to analytically continuing the result to the region of interest.

In the cubic model we are interested in $0<\z-d/5\ll 1$.
In order to extract the coefficient $b$ of equation \eqref{eq:bt} we take a derivative with respect to $\omega_0^2$, and set $\omega_0=\mu^\z$ and $\zeta=\frac{d+\eps}{5}$. We find:
\begin{equation}
\partial_{\omega_0^2} \mathcal{B}(\mu^\z,0)=- \mu^{-\eps}\frac{2^{-d-3} \pi ^{-\frac{d}{2}} }{\epsilon  \Gamma \left(\frac{d}{2}\right)}+O(1) \; ,  
\end{equation}
and therefore
\begin{equation}
    b=\frac{2^{-d-3} \pi ^{-\frac{d}{2}} }{  \Gamma \left(\frac{d}{2}\right)} \; .
\end{equation}

In the quartic model, we are instead interested in $0<\z-d/3\ll 1$, which lies in the range of convergence.
In this case, in order to extract the coefficient $B$ of equation \eqref{eq:BM-phi4} we do not take any derivative, and we directly set $\omega_0=\mu^\z$ and $\zeta=\frac{d+\eps}{3}$. We find:
\begin{equation}
 \mathcal{B}(\mu^\z,0)= \mu^{-\eps} \frac{2^{-d-1} \pi ^{-\frac{d}{2}}}{\epsilon  \Gamma \left(\frac{d}{2}\right)}+O(1) \; ,  
\end{equation}
and therefore
\begin{equation} \label{eq:B}
    B = \frac{2^{-d-1} \pi ^{-\frac{d}{2}}}{  \Gamma \left(\frac{d}{2}\right)} = 4 b \; .
\end{equation}

\subsection{The triangle integral, $\mathcal{T}$}
The triangle integral at the subtraction point writes:
\begin{equation}
    \mathcal{T}(\mu^\zeta,0;-\mu^\zeta,0)= \int \frac{d\omega d^d q}{(2\pi)^{d+1}} \; \frac{1}{ [ \omega^2+(q^2)^{\zeta} ]^2 } \; \frac{1}{ (\omega+\mu^\zeta)^2+ (q^2)^{\zeta} } \ .
\end{equation}
Using again the residue theorem we find: 
\begin{equation}
    \mathcal{T}(\mu^\zeta,0;-\mu^\zeta,0)=\int \frac{d^d q}{(2\pi)^d} \frac{q^{-3 \zeta } \left(12 q^{2 \zeta }+\mu^{2\zeta}\right)}{4 \left(4 q^{2 \zeta }+\mu^{2\zeta}\right)^2}=\frac{\pi ^{1-\frac{d}{2}} 2^{1-\frac{d (\zeta +1)}{\zeta }} (d-2 \zeta ) \mu^{d-5\zeta} }{\zeta ^2 \Gamma \left(\frac{d}{2}\right)\cos \left(\frac{\pi  d}{2 \zeta }\right)} \ ,
\end{equation}
where the spatial momentum integral is convergent in the interval $\z\in(d/5,d/3)$, which includes $\z=\f{d+\eps}{5}$.
For small $\eps$ we find:
\begin{equation}
        \mathcal{T}(\mu^\zeta,0;-\mu^\zeta,0)=\mu^{-\eps} \, 3\,\frac{2^{-d-3} \pi ^{-\frac{d}{2}}}{\epsilon  \Gamma \left(\frac{d}{2}\right)} +O(1) \; ,
\end{equation}
and therefore, the coefficient $t$ of equation \eqref{eq:bt} is:
\begin{equation}
        t=3\,\frac{ 2^{-d-3} \pi ^{-\frac{d}{2}}}{  \Gamma \left(\frac{d}{2}\right)} =3b \; .
\end{equation}

We can also verify that the above result is independent of the choice of $c$ in \eqref{eq:ren-cond}.
For general $c$ we have:
\begin{equation}
    \mathcal{T}(\mu^\z,0;-(1+c)\mu^\z,0)=\int \frac{d\omega d^d q}{(2\pi)^{d+1}} \; \frac{1}{  \omega^2+(q^2)^{\zeta}  } \; \frac{1}{ (\omega+\mu^\zeta)^2+ (q^2)^{\zeta} } \frac{1}{ (\omega-c \mu^\zeta)^2+ (q^2)^{\zeta} } \ .
\end{equation}
The integration of the frequency can be performed again by residue theorem, and the remaining integral in spherical coordinates is 
\begin{equation}
    \mathcal{T}(\mu^\z,0;-(1+c)\mu^\z,0)=\frac{2^{1-d} \pi ^{-\frac{d}{2}}}{\Gamma \left(\frac{d}{2}\right)} \int_0^{+\infty} d q \ q^{d-1} \frac{q^{-\zeta } (12 q^{2 \zeta }+(1+c+c^2) \mu^{2\zeta} )}{(4 q^{2 \zeta }+\mu^{2\zeta})(4 q^{2 \zeta }+c^2 \mu^{2\zeta}) (4 q^{2 \zeta }+(1+c)^2\mu^{2\zeta})} \; ,
\end{equation}
which we could not evaluate explicitly. However, we can extract the $1/\eps$ pole by the following reasoning.
The integral is convergent for $\z\in(d/5,d)$. At $\z=\frac{d}{5}$, it is logarithmically divergent, and as it does not contain sub-divergences, in analytic regularization it diverges as a simple pole in $\epsilon$. Moreover, the coefficient of such pole does not depend on the infrared regulator. Since the ultraviolet divergence that generates the $1/\epsilon$ pole arises at large momenta, we can set $\mu=0$ in the integrand and compute the integral with a sharp cut-off instead to extract the universal coefficient of the $\epsilon$ pole. We find as before:
\begin{equation}
    \mathcal{T}(\mu^\z,0;-(1+c)\mu^\z,0)=\mu^{-\eps}\, 3\,\frac{ 2^{-d-3} \pi ^{-\frac{d}{2}}}{\epsilon  \Gamma \left(\frac{d}{2}\right)}+O(1) \; .
\end{equation}

\subsection{The melon integral, $\mathcal{M}$}

The melon integral writes:
\begin{equation}
    \mathcal{M}(\omega_0,0) = \int \frac{d^dq_1 \, d\omega_1}{(2\pi)^{d+1} }  \int \frac{d^dq_2 \, d\omega_2}{(2\pi)^{d+1} }  \,\frac{1}{ \omega_1^2 + (q_1^2)^\zeta } \, 
\,\frac{1}{ \omega_2^2 + (q_2^2)^\zeta } \, 
\frac{1}{ (\omega_0+\omega_1+\omega_2)^2 + 
[( q_1+q_2)^2 ]^\zeta } \;,
\end{equation}
which becomes, using the Schwinger parametrization:
\begin{equation}
\begin{split}
    \mathcal{M}(\omega_0,0) &= \int \frac{d^dq_1 \, d\omega_1}{(2\pi)^{d+1} }  \int \frac{d^dq_2 \, d\omega_2}{(2\pi)^{d+1} }  \, \int_0^{+\infty} d\a_1 d\a_2 d\a_3 \\
    &\qquad\qquad \times  \,e^{-\a_1 (\omega_1^2 + (q_1^2)^\zeta ) -\a_2 (\omega_2^2 + (q_2^2)^\zeta )
-\a_3 ( (\omega_0+\omega_1+\omega_2)^2 + 
[( q_1+q_2)^2 ]^\zeta ) } \\
&= \int \frac{d^dq_1 }{(2\pi)^{d} }  \int \frac{d^dq_2}{(2\pi)^{d} }  \, \int_0^{+\infty} d\a_1 d\a_2 d\a_3 \\
    &\qquad\qquad \times  \,e^{-\a_1  (q_1^2)^\zeta  -\a_2  (q_2^2)^\zeta -\a_3 [( q_1+q_2)^2 ]^\zeta  } 
\frac{ e^{-\frac{\a_1 \a_2 \a_3 \omega_0^2}{\a_1 \a_2 + \a_1 \a_3+\a_2 \a_3}}}{4\pi\sqrt{\a_1 \a_2 + \a_1 \a_3+\a_2 \a_3}} \;.
\end{split}
\end{equation}
Next, taking the derivative with respect to $\omega_0^2$ and setting the latter to zero, we obtain
\begin{equation}
\begin{split}
    \p_{\omega_0^2}\mathcal{M}(0,0) 
&= - \int_{|q_1|\geq \m} \frac{d^dq_1 }{(2\pi)^{d} }  \int_{|q_2|\geq \m} \frac{d^dq_2}{(2\pi)^{d} }  \, \int_0^{+\infty} d\a_1 d\a_2 d\a_3 \\
    &\qquad\qquad \times   \,e^{-\a_1  (q_1^2)^\zeta  -\a_2  (q_2^2)^\zeta -\a_3 [( q_1+q_2)^2 ]^\zeta  } 
\frac{ \a_1 \a_2 \a_3}{4\pi (\a_1 \a_2 + \a_1 \a_3+\a_2 \a_3)^{3/2}} \;.
\end{split}
\end{equation}
The subscript on the momentum integrals signals that, as we have set $\omega_0=0$, the integral is infrared divergent and needs a cutoff $\m$.
As we are only interested in the leading ultraviolet divergence, the precise form of the infrared regulator is irrelevant.

In the sector $\a_3>\a_1,\a_2$, we make the redefinition $\a_i=t_i \a_3$ for $i=1,2$, and integrate over $\a_3$. Taking into account that we have 3 sectors, we obtain:
\begin{equation}
\begin{split}
    \p_{\omega_0^2}\mathcal{M}(0,0) 
&= - \f{3}{2\pi} \int_0^1 dt_1 dt_2   
\frac{ t_1 t_2 }{(t_1 +t_2 +t_1 t_2)^{3/2}}  \\
    &\qquad\qquad \times \,
\int_{|q_1|\geq \m} \frac{d^dq_1 }{(2\pi)^{d} }  \int_{|q_2|\geq \m} \frac{d^dq_2}{(2\pi)^{d} } 
\frac{1}{({t_1  (q_1^2)^\zeta  +t_2  (q_2^2)^\zeta + [( q_1+q_2)^2 ]^\zeta  })^{3}} \;.
\end{split}
\end{equation}

In order to extract the residue of $1/\eps$ pole we use the following strategy.
For $d>1$, we go to spherical coordinates and integrate out the angular coordinates, except the angle between $q_1$ and $q_2$, which we denote $\theta$.
Then, we define a vector with the moduli of $q_1$ and $q_2$ as components $q=(|q_1|,|q_2|)$, and we switch to polar coordinates $q\rightarrow (p,\varphi)$, with $p=|q|$ and $\varphi\in(0,\pi/2)$. We get:
\begin{equation}
\begin{split}
    \p_{\omega_0^2}\mathcal{M}(0,0) 
&= - \frac{3 S_{d-1} S_{d-2}}{(2\pi)^{2d+1}} \int_0^1 dt_1 dt_2   
\frac{ t_1 t_2 }{ (t_1 +t_2 +t_1 t_2)^{3/2}} \int_{\sqrt{2}\m}^\infty dp \frac{p^{2d-1}}{p^{6\z}} \\ & \quad\times\, \int_0^{\pi}d\theta \int_{0}^{\pi/2}d\varphi  \frac{\sin ^{d-2}(\theta ) \sin ^{d-1}(\varphi ) \cos ^{d-1}(\varphi )}{\left(t_1 \left(\cos ^2(\varphi )\right)^{\zeta }+t_2 \left(\sin ^2(\varphi )\right)^{\zeta }+\left[1+2 \cos (\theta ) \sin (\varphi ) \cos (\varphi )\right]^{\zeta }\right)^3} \; ,
\end{split}
\end{equation}
where $S_{n}$ is the volume of the $n$-sphere.
 
The integral over $p$ is convergent for $\z= \frac{d+\epsilon}{3}>d/3$ and it has a $1/\eps$ pole, $\int_{\sqrt{2}\m}^\infty  \; p^{-1 -2\epsilon} = \frac{\mu^{-2\epsilon}}{2^{1+\eps} \, \epsilon}$. It follows that the coefficient $M$ in Eq.~\eqref{eq:BM-phi4} is:
\begin{equation} \label{eq:M_2}
\begin{split}
    M &= \frac{3 S_{d-1} S_{d-2}}{2 (2\pi)^{2d+1}} \int_0^1 dt_1 dt_2   
\frac{ t_1 t_2 }{(t_1 +t_2 +t_1 t_2)^{3/2}}  \\ & \quad\times\,
\int_0^{\pi}d\theta \int_{0}^{\pi/2}d\varphi  \frac{\sin ^{d-2}(\theta ) \sin ^{d-1}(\varphi ) \cos ^{d-1}(\varphi )}{\left(t_1 \left(\cos ^2(\varphi )\right)^{d/3 }+t_2 \left(\sin ^2(\varphi )\right)^{d/3 }+\left[1+2 \cos (\theta ) \sin (\varphi ) \cos (\varphi )\right]^{d/3 }\right)^3} \; ,
\end{split}
\end{equation}
and the remaining integrals are convergent and can be evaluated numerically.

For $d=1$, there is no angular integration for the two momenta, and performing directly the change of variables $(q_1,q_2)\rightarrow (p,\varphi)$, where now $\varphi\in(0,2\pi)$, we get:
\begin{equation} \label{eq:M_1}
\begin{split}
    M &= \frac{3 }{2 (2\pi)^{3}} \int_0^1 dt_1 dt_2   
\frac{ t_1 t_2 }{(t_1 +t_2 +t_1 t_2)^{3/2}}  \\ & \quad\times\,
\int_{0}^{2\pi}d\varphi  \frac{1}{\left(t_1 \left(\cos ^2(\varphi )\right)^{1/3}+t_2 \left(\sin ^2(\varphi )\right)^{1/3 }+\left[1+2  \sin (\varphi ) \cos (\varphi )\right]^{1/3 }\right)^3} \; .
\end{split}
\end{equation}
A numerical evaluation with 1\% precision gives:
\begin{equation}
    M\simeq\begin{cases}
        3.19 \times 10^{-4} &\; \text{for } d=2 \\
        2.66 \times 10^{-3} &\; \text{for } d=1
    \end{cases} \; .
\end{equation}

\bibliographystyle{JHEP}
\bibliography{Refs,Refs_new} 
\addcontentsline{toc}{section}{References}

\end{document}